\pdfoutput=1

\PassOptionsToPackage{x11names,svgnames}{xcolor}

\documentclass{SciPost}
\hypersetup{
    colorlinks,
    linkcolor={red!50!black},
    citecolor={blue!50!black},
    urlcolor={blue!80!black}
}

\usepackage[bitstream-charter]{mathdesign}
\DeclareSymbolFont{usualmathcal}{OMS}{cmsy}{m}{n}
\DeclareSymbolFontAlphabet{\mathcal}{usualmathcal}

\fancypagestyle{SPstyle}{
\fancyhf{}
\lhead{\colorbox{scipostblue}{\bf \color{white} ~SciPost Physics Codebases }}
\rhead{{\bf \color{scipostdeepblue} ~Submission }}

\fancyfoot[C]{\textbf{\thepage}}
}

\usepackage{doi}

\usepackage{xspace}
\usepackage{graphicx}
\usepackage{afterpage}
\usepackage{lineno}
\usepackage{tikz}
\usetikzlibrary{arrows,shapes,backgrounds,fit,positioning,calc}
\definecolor{irn-bru}{RGB}{239,129,0}

\usepackage{subfig}
\usepackage{microtype}
\makeatletter
\g@addto@macro\bfseries{\boldmath}
\makeatother

\usepackage{xpatch}
\usepackage{xparse}

\usepackage{listings}
\lstMakeShortInline|

\usepackage{todonotes}
\makeatletter
\xpretocmd{\todo}{\@bsphack}{}{}
\xapptocmd{\todo}{\@esphack}{}{}
\makeatother

\usepackage{siunitx}
\DeclareSIUnit{\electronvolt}{\text{e\kern-0.15ex V}}
\DeclareSIUnit{\eV}{\electronvolt}
\DeclareSIUnit{\MeV}{\mega\eV}
\DeclareSIUnit{\GeV}{\giga\eV}
\DeclareSIUnit{\TeV}{\tera\kern-0.1ex\eV}
\DeclareSIUnit{\ifb}{\femto\barn\tothe{-1}}

\newcommand{\cxx}{\text{C\protect\raisebox{0.15ex}{\scalebox{0.8}{++}}}\xspace}

\NewDocumentCommand{\rivet}{o}{\text{R\protect\scalebox{0.8}{IVET}\IfNoValueTF{#1}{}{\,#1}}\xspace}
\NewDocumentCommand{\contur}{o}{\text{C\protect\scalebox{0.8}{ONTUR}\IfNoValueTF{#1}{}{\,#1}}\xspace}
\NewDocumentCommand{\yoda}{o}{\text{Y\protect\scalebox{0.8}{ODA}\IfNoValueTF{#1}{}{\,#1}}\xspace}
\newcommand{\hepdata}{HEPD\protect\scalebox{0.8}{ATA}\xspace}

\makeatletter
\let\@oldto\to
\renewcommand{\to}{\ensuremath{\@oldto}\xspace}
\renewcommand{\to}{\ensuremath{\@oldto}\xspace}

\makeatother

\renewcommand{\vec}[1]{\mathbf{#1}}

\NewDocumentCommand{\varcoord}{o}{\ensuremath{\IfNoValueTF{#1}{\vec{\omega}}{\omega^{(#1)}}}\xspace}

\NewDocumentCommand{\bincoord}{o}{\ensuremath{\IfNoValueTF{#1}{\vec{\theta}}{\theta^{(#1)}}}\xspace}

\NewDocumentCommand{\unbincoord}{o}{\ensuremath{\IfNoValueTF{#1}{y}{y}}\xspace} 

\hypersetup{hidelinks,
  pdfauthor={Christian Bierlich, Andy Buckley, Jonathan Butterworth, Christian Gutschow, Leif Lonnblad, Tomasz Procter, Peter Richardson, Yoran Yeh}
  pdftitle={Robust Independent Validation of Experiment and Theory: Rivet version 4 release note}
}

\begin{document}
\pagestyle{SPstyle}

\begin{center}
  \color{scipostdeepblue}
  \Large
  \textbf{Robust Independent Validation of Experiment and Theory: \rivet version 4 release note}
\end{center}

\begin{center}
  C.~Bierlich$^1$,
  A.~Buckley$^2$,
  J.~Butterworth$^3$,
  C.~G\"utschow$^{3,4,*}$,
  L.~L\"onnblad$^1$,\\
  T.~Procter$^2$,
  P.~Richardson$^6$,
  Y.~Yeh$^3$,
\end{center}

\begin{center}
  \itshape
  $^1$ Department of Physics, Lund University\\ Professorsgatan 1, 223~63 Lund, Sweden\\
  $^2$ School of Physics \& Astronomy, University of Glasgow,\\ University~Place, G12~8QQ, Glasgow, UK\\
  $^3$ Department of Physics \& Astronomy, University College London,\\ Gower~Street, WC1E~6BT, London, UK\\
  $^4$ Centre for Advanced Research Computing, University College London,\\ Gower~Street, London, WC1E~6BT, UK\\
  $^5$ Institute for Particle Physics Phenomenology, Durham University\\ South Road, Durham, DH1~3LE, UK\\
\end{center}

\begin{center}
\today
\end{center}


\section*{Abstract}
The \rivet toolkit is the primary mechanism for phenomenological
preservation of collider-physics measurements, containing both a
computational core and API for analysis implementation, and a large
collection of more than a thousand preserved analyses.  In this note
we summarise the main changes in the new \rivet[4] major release
series. These include a major generalisation and more semantically
coherent model for histograms and related data objects, a thorough
clean-up of inelegant and legacy observable-computation tools, and new
systems for extended analysis-data, incorporation of preserved
machine-learning models, and serialization for high-performance
computing applications. Where these changes introduce
backward-incompatible interface changes, existing analyses have
been updated and indications are given on how
to update new analysis routines and workflows.

\clearpage

\begin{centering}
  \small
  \textbf{Analysis-routine authors since \rivet[3]}\\
  M.~I.~Abdulhamid,
M.~Alvarez,
S.~H.~An,
M.~Azarkin,
L.~I.~E.~Banos,
S.~Bansal,
F.~Barreiro,
A.~Basan,
E.~Berti,
B.~Bilin
% ...

\end{centering}

\section{Introduction}
\label{sec:intro}

The \rivet toolkit~\cite{Bierlich:2019rhm,Buckley:2010ar} is the
primary mechanism for phenomenological preservation of
collider-physics measurement analyses, most prominently but by no
means exclusively at the LHC. \rivet consists of a computational core
for theoretically robust reproduction of experimental data-processing,
including a configurable system for smearing-based detector emulation,
and a large library of more than \num{1500} preserved analyses.

The \rivet workflow is centred on explicit runs of simulated collider
events, produced with external event-generator tools and fed to \rivet
in the standard HepMC\,3 event-record~\cite{Buckley:2019xhk}
structure. This input can either be in any of the serialized data
formats supported by HepMC or, reducing I/O performance overheads, via
in-memory object exchange; it is important that the events implement
the standard PDG particle-ID coding
scheme~\cite{ParticleDataGroup:2022pth}, use HepMC's standard for
particle status codes, and for full functionality follow the MCnet
event-weights standard~\cite{Bothmann:2022pwf}.

The role of \rivet is then to analyse each event, reconstruct
quantities equivalent to those used by the experimental analyses for
event selection and data processing (e.g. cuts, histogramming, or as
machine-learning inputs) and to output datasets directly comparable
with the public data measurements. This is made more expressive and
efficient by \rivet's library of standard physics objects and
observable-calculator tools, many of which are coded as
\emph{projections} that automatically cache and reuse their
calculations when appropriate.  The analysis codes are written in
physics-focused C++, and loaded at runtime from \emph{plugin}
libraries. The standard set of analysis plugins is installed
automatically along with the core framework, including reference
datasets from the \hepdata repository~\cite{Maguire:2017ypu}, meaning
that a user in possession of a HepMC generator or dataset can compare
its predictions to perhaps hundreds of relevant data analyses with
minimal effort.

\rivet[4] is a new major release series, significantly refining and
extending the treatment of data objects, providing mechanisms for
efficient parallel running on MPI clusters and for inclusion of
machine-learning analysis components, and improving the
observable-computation tools. These changes have necessitated some
backward-incompatible changes to the programming interface (or ``API'')
exposed to analysis-author users, as well as extensions to the v3
functionality. In this note we summarise the major developments, as
well as directing users to what needs to be done to adapt to the API
changes.

\section{Histogramming}
\label{sec:histo}

\rivet's histogramming system has been reworked to support the new
major release series of the \yoda[2]~\cite{Buckley:2023xqh}
statistical data-analysis library. This has significantly simplified
\rivet's internal \yoda interface, while extending histogramming
support to arbitrary histogram and profile dimensions based on modern
\cxx template techniques. \rivet retains the ability to coherently
synchronise correlated NLO subevents into histograms, in addition to
handling arbitrary numbers of systematic-variation weight streams;
this has similarly been generalised to arbitrary histogram dimensions.

\rivet[4] fully embraces the concept of \emph{live} and \emph{inert}
data objects introduced in \yoda[2].  All of the experimental
reference data shipped with \rivet is now represented in terms of
\texttt{Estimate}-type objects, which lend themselves better to
measurement data than the \texttt{Scatter}-type objects. This
addresses a long-standing issue where intrinsically discrete
observables such as multiplicity distributions had to be represented
with dummy bin-widths around integer values: while \texttt{Scatter}s
had to be heuristically interpreted as implying binnings, the
\texttt{Estimate}s implement continuous or discrete data-binning as
appropriate to the data. The new object types are derived from the
same underlying \texttt{BinnedStorage} formalism championed by
\yoda[2].

In the same spirit, any \emph{live} types of analysis objects left at
the end of the \texttt{finalize()} stage will be automatically
converted to \emph{inert} estimate types, usually a
\texttt{BinnedEstimate1D}. This ensures full type consistency and
direct comparability between the objects produced in a \rivet run and
the reference data.  Pre-\texttt{finalize} copies of analysis of all
analysis data-objects, whose paths are prefixed with \texttt{/RAW},
remain live types, allowing re-entrant finalization --
i.e.~statistically exact merging or extension of MC runs -- via the
\texttt{rivet-merge} command-line utility. Custom conversions to inert
form can be explicitly performed in \texttt{finalize()}, e.g.~by using
the \texttt{barchart()} function to bypass the default scaling by
histogram bin widths.

The confusing \texttt{BinnedHistogram} type has been replaced by a new
\texttt{HistoGroup} class, which better reflects that this object
is essentially a binning of histograms.  While the new type has the
same aim and basic design, it is not implemented as a disjoint
standalone class, but rather exploits the fact that \yoda[2]'s new
\texttt{BinnedStorage} class is agnostic about its templated bin
content, and could itself derive from \texttt{BinnedStorage}, thereby
realising a literal histogram-of-histograms.

As well as the statistical core of the generalised \yoda live and
inert histogramming, interfaced with \rivet's counter-event and
weight-stream coordination, \rivet[4] features a new histogram
rendering system. This is built on \yoda[2]'s \texttt{matplotlib}
plotting system, and replaces the venerable \text{make-plots} script
based on \LaTeX-\texttt{pstricks}. The latter produced high-quality
output, but was also inflexible and sometimes unstable due to the
antiquated backend; the new version reproduces the previous style
almost exactly via styling additions to the \yoda defaults, within the
more modern and supported \texttt{matplotlib} ecosystems, and
maintains the publication-friendly model of also writing intermediate
script files for customisation. Finer details of the cosmetic layout
are defined in dedicated \texttt{plot} files, so as to not distract
from the physics in the main analysis plugins. Existing style files
are still supported by the new plotting infrastructure, but newer
versions based on the widely used YAML format are now also accepted.

\section{Projection streamlining}

Many of the existing \texttt{enum}s have been renamed and often
decoupled from specific projection scopes to present a more uniform,
predictable, and less deeply nested interface for projection
configuration. Projection constructors and methods with unclear
\texttt{bool} arguments have been converted to use these more
``self-documenting'' \texttt{enum}s. Scoped \texttt{enum} classes are
now used as standard, enforcing type consistency while improving
readability and self-documentation, and are defined outside the
projection scopes to reduce the amount of scope chaining.

This simple but widespread change eliminates the previous inconsistent mix
of two- and three-layered \texttt{enum}s, for instance from both the
\texttt{JetAlg} and \texttt{FastJets} scopes when configuring jet
finders. Examples include the relabelling of
\texttt{FastJets::Algo::KT} to \texttt{JetAlg::KT},
of
\texttt{JetAlg::Muons::NONE} to \texttt{JetMuons::NONE},
or from
\texttt{JetAlg::Invisibles::DECAY} to \texttt{JetInvisibles::DECAY}.
Similarly,
the boolean arguments triggering whether or not tau and muon decay
particles are treated as prompt have been replaced by
\texttt{TauDecaysAs} or \texttt{MuDecaysAs} \texttt{enum} classes,
respectively.

The \texttt{DressedLeptons} projection has been renamed
\texttt{LeptonFinder}, bringing it more in line with the existing
\texttt{ParticleFinder} and \texttt{JetFinder} classes while being
agnostic about the dressing logic used to reconstruction the
leptons. Note that the old \texttt{DressedLeptons} class name is now
an alias for a \texttt{vector} of \texttt{DressedLepton} objects,
which again streamlines the type-naming semantics with respect to the
existing \texttt{Particles} and \texttt{Jets} aliases.

The \texttt{ZFinder} has been similarly canonically renamed to
\texttt{DileptonFinder}, better reflecting both the mass-generality
and lepton-specificity of its approach. The constructor arguments have
been simplified and the previously implicit target mass is now a
explicitly passed constructor argument. The old and often
misunderstood \texttt{trackPhotons} argument has been removed -- the
same strategy can be realised by setting the \texttt{dR} argument to
\texttt{-1}.

The \texttt{WFinder} has been removed entirely as it had too many
analysis options, reflecting the need for analysis-specific heuristics
to compensate for the imperfect information induced by the invisible
neutrino. It is preferable to implement the analysis-specific
selection cuts manually in the analysis, e.g.~using dressed leptons
and missing tranverse energy explicitly via the new
\texttt{closestMatchIndex()} metafunction used with the \texttt{mass()}
or \texttt{mT()} unbound functions as appropriate. Applying this
migration to the standard analysis collection was found to
significantly improve the self-documentation of analysis code,
clarifying previously obscure model-dependent assumptions woven into
the measurement data.

Finally the jet-smearing system via the \texttt{SmearedJets}
projection now accepts an ordered list of smearing and efficiency
functions via a \cxx parameter pack.  This has required a re-ordering
of parameters in a slightly non-optimal fashion, but was considered an
overall preferable solution, especially given that this feature is a
somewhat niche requirement for cases where multiple distinct detector
effects need to be applied in a specific order.

A FAQ document supplying detailed advice and mechanics for migration of
\rivet[3] analysis routines to the v4 API is provided on the \rivet
code repository, and linked from the website documentation.
New users will find that the majority of presentation in the \rivet[3]
paper~\cite{Bierlich:2019rhm} remains valid for \rivet[4], hence the
brevity of this release note, but the best resources for practical
familiarisation and problem-solving are the tutorial pages ---
similarly found in the code repository --- in addition to resources on
from live tutorial events on the \rivet website.
As always, any analyses supplied to the team for inclusion in the standard
collection are maintained and migrated by the \rivet team, but support
is available to all who need help with updating their codes to the new
release series.

\section{Additional interface improvements}

Older existing analyses previously named with a \textsc{Spires}-based
identifier have now been renamed to match newer analyses whose name is
already based on their \textsc{Inspire} code.  Previous aliasing to
allow loading via the \textsc{Inspire} name has now been switched such
that the non-canonical alias is the \textsc{Spires} name. These
fallback aliases will eventually be removed.

Building on the similar feature in \yoda[2], this release of \rivet
introduces a serialization mechanism for the core
\texttt{AnalysisHandler} objects, so their metadata and list of
analysis objects can be efficiently rendered into a stream of
\texttt{double}-valued numbers, and then reconstructed.  While of
potentially general usefulness, this mechanism allows synchronisation
of many different parallel \rivet instances in an MPI-based
high-performance computing cluster, without requiring very expensive
access to the filesystem. This mechanism has been demonstrated as an
effective component of combined high-precision event generation,
showering, and analysis on leading
HPC clusters~\cite{Hoche:2019flt,Bothmann:2022thx,Bothmann:2023ozs}.
Running with multiple simultaneous \texttt{AnalysisHandler} objects
across many threads is also now possible, with ability to merge the
handlers at the end of the run.

\rivet[4] adds functionality for storing and loading analysis-specific
structured data in the widely used \texttt{HDF5} format, making
\texttt{HDF5} as well as the lightweight \texttt{HighFive} I/O layer
strict dependencies in the process. The (new) canonical analysis name
is used as a filename prefix to ensure specificity to the associated
analysis routine, and grouping of all files for each analysis in
directory listings.

A specialisation of this analysis-data mechanism has been developed to
assist import of machine-learning (ML) models, in particular deep
neural-net and graph neural-net architectures, from ONNX serialisation
files stored specific to the analysis. While some question marks
remain over the long-term stability of ML models, e.g.~with respect to
evolution of their implementing frameworks, ONNX has been identified
as the current best option for ML preservation~\cite{Araz:2023mda}.
This option is implemented as a header file only, so the
ONNX~Runtime~\cite{onnxruntime} library is only an optional dependency
for \rivet. An example usage is given below:\\[1ex]
\null\quad\texttt{\_nn = getONNX(name());}\\
\null\quad\texttt{...}\\
\null\quad\texttt{vector<float> nn\_outputs = \_nn->compute(nn\_input);}\\[1ex]
The first line, likely in the analysis \texttt{init()} function,
retrieves a single ONNX object registered under the analysis
routine's own name; if there were multiple analyses, they would be
registered with different suffixes following the analysis
name.\footnote{ Having to manually specify the name is a limitation of
the current header-level decoupling of these functions from the \rivet
\texttt{Analysis} class hierarchy.} The second call is the use of the
neural network, generally returning a vector of numeric values when
passed a vector of event features. Further helper functions are
available for e.g.~retrieval of ONNX metadata.

If desired, tighter ONNX~Runtime integration can be
enabled at build-time: when enabled, its compiler and linker flags are
automatically propagated to the \texttt{rivet-build} script, and the
(currently small) collection of ONNX-using analysis routines will be
built and installed. Subject to the success of ONNX-preservation
initiatives in collider-physics collaborations, this dependency may
become mandatory in a future release in the \rivet[4] series.

For simplification of interfacing code, and to reflect its uptake and
development status as the new standard particle-level event format in
the high-energy physics community, support has been removed for the
HepMC\,2 API and library structure: the HepMC\,2 file format can still
be read, via the I/O routines of the HepMC\,3 library.

\section{Conclusions and outlook}
\label{sec:concl}

The new release series of the \rivet toolkit introduces a number of
API-breaking changes, in order to better support more coherent
workflows and ensure more self-documenting analysis
routines. Additional generalisation provide official mechanisms to
load complex auxiliary analysis data, including ONNX machine-learning
models, and to use \rivet efficiently in HPC environments. The overall
effect of these collective changes is a very positive evolution in
\rivet's interface and capabilities, cementing the characteristics
that have made it a successful tool for collider-physics analysis
preservation. As with all previous versions, \rivet[4] is available
for installation from source and as Docker images, as detailed at
\url{https://rivet.hepforge.org/}.

\section{Acknowledgements}
The authors thank the Marie Sklodowska-Curie Innovative Training
Network MCnetITN3 (grant agreement no. 722104) for funding and
providing the scope for discussion and collaboration toward this
work. AB and CG acknowledge funding via the STFC experimental
Consolidated Grants programme (grant numbers %
ST/S000887/1 \& 
ST/W000520/1 
and ST/S000666/1) \& 
ST/W00058X/1), 
and the SWIFT-HEP project (grant numbers
ST/V002562/1 
and ST/V002627/1). 
YY thanks the Spreadbury Fund and the UCL Impact scheme for PhD studentship funding.
Many thanks to Markus Seidel, Alex Grecu, and Antonin Maire for
support as additional LHC experiment contacts.
Our thanks also to all the many user-contributors whose inputs and
support have enabled \rivet to grow and evolve, and who provide such a
welcoming user community, in particular
Enrico Bothmann,
Christian Holm Christensen,
Stefan H{\"o}che,
Dmitry Kalinkin,
Stefan Kiebacher,
Max Knobbe,
Alexander Puck Neuwirth,
Marek Sch{\"o}nherr,
Andrii Verbytskyi,
and
James Whitehead.

\bibliography{rivet4.bib}

\end{document}